\documentclass[aps,prd,amsmath,twocolumn,superscriptaddress]{revtex4}

\usepackage{hyperref}
\usepackage{ifthen}
\usepackage{amsmath}
\usepackage{amssymb}
\usepackage{color}
\usepackage{graphicx}
\usepackage{xfrac}

\definecolor{verde}{rgb}{0,0.5,0}

\def\be{\begin{equation}}
\def\ee{\end{equation}}
\def\bea{\begin{eqnarray}}
\def\eea{\end{eqnarray}}
\def\be{\begin{equation}}
\def\ee{\end{equation}}
\def\ba{\begin{align}}
\def\ea{\end{align}}
\def\p{\partial}
\def\noi{\noindent}

\renewcommand\]{\right]}

\newcommand\lsim{\mathrel{\rlap{\lower4pt\hbox{\hskip0.5pt$\sim$}}
    \raise1pt\hbox{$<$}}}
\newcommand\gsim{\mathrel{\rlap{\lower4pt\hbox{\hskip0.5pt$\sim$}}
    \raise1pt\hbox{$>$}}}

\begin{document}

%
\input epsf
\renewcommand{\topfraction}{0.99}
\renewcommand{\bottomfraction}{0.99}

\title{Non-linear Fields in Generalized Cosmologies}
\author{Matteo Fasiello$^1$ and Zvonimir Vlah}
\affiliation{Stanford Institute for Theoretical Physics and Department of Physics, Stanford University, Stanford, CA 94306 }
\affiliation{Kavli Institute for Particle Astrophysics and Cosmology,
Stanford University and SLAC, Menlo Park, CA 94025 }

\begin{abstract}
\noindent 
The perturbative approach to structure formation has recently received a lot of attention in the literature. In such setups the final predictions for observables like the power spectrum is often derived under additional approximations such as a simplified time dependence. Here we provide all-order perturbative integral solutions for density and velocity fields in generalized cosmologies, with a direct application to clustering quintessence.  We go beyond the standard results based on extending the EdS-like approximations. As an illustrative example, we apply our findings to the calculation of the one-loop power spectrum of density and momentum fields. 
We find corrections close to $1\%$ in the mildly non-linear regime of $\Lambda$CDM cosmologies for the density power spectrum, while in the case of the density-momentum power spectrum effects can reach up to 1.5\% for $k\sim 0.2h/$Mpc. 
\end{abstract}


\maketitle
\noindent
\noindent
\section {Introduction}

A detailed understanding of the-nonlinear formation of structure in 
the Universe is of  paramount importance for cosmology. 
Several approaches have been developed to tackle the dynamics at these scales. State of the art N-body simulations give reliable answers at 1\% level in the power spectrum 
up to $k \sim 1$ h/Mpc \cite{Heitmann:2008eq}, but these can be expensive to run and may not be fully convergent using typical 
current generation box size and resolution. It must also be noted that the accuracy an reliability of N-body simulations is a fast and ever-improving process.

Clustering statistics is used to extract information on our Universe, and simulations do not necessarily provide the deepest insight into how to  identify the most useful information content and to optimally extract it from  data.  A complementary approach to the non-linear scales is based upon extending the reach of perturbation theory towards the quasi-linear regime. 
One practical advantage of the perturbative approach relies in the faster evaluation of observables for a given set of cosmological parameters
(see e.g. \cite{Schmittfull:2016jsw, McEwen:2016fjn} for recent improvements implementing FFT's) used in the analysis of cosmological measurements.
Perturbation theory also offers the analytical handle which is best suited to probe the physical principles underlying the data.

On this basis, a lot of effort has been put towards computing the statistical properties of the density distribution. In an non-exhaustive list, we mention here the Eulerian perturbation theory framework \cite{Bouchet:1995ez, Bernardeau:2001qr, Carlson:2009it, Bernardeau:2013oda}), renormalized perturbation theory in the form of RPT (renormalized perturbation theory) \cite{Crocce:2005xy,Crocce:2005xz}, \cite{Matarrese:2007wc}, TRG (time renormalization group)\cite{Pietroni:2008jx}, TSPT (time-sliced perturbation theory) \cite{Blas:2015qsi}  and the EFTofLSS (effective field theory of large scale structure) program \cite{Baumann:2010tm,Carrasco:2012cv,Carrasco:2013sva,Carrasco:2013mua,Bertolini:2016bmt,Pajer:2013jj}. The Lagrangian approach has also been successfully implemented \cite{Matsubara:2008wx, Matsubara:2007wj, Carlson:2012bu, Porto:2013qua, Vlah:2014nta, Vlah:2015sea}. 

In this paper we give a thorough derivation of the solutions for density fluctuation to all orders in perturbation theory accounting for a non-trivial 
non-factorizable time-dependence which maybe easily expanded to more general cosmologies. 
We extend the standardly used approximations (with notable exceptions, e.g. \cite{Takahashi:2008yk,Carrasco:2012cv,Rampf:2015mza}) where it is assumed that the gravity kernels are time-independent.  We further use these results to calculate the 1-loop density power spectrum as well as density-momentum power spectrum.

We stress here the realm of validity of our analysis: our starting point is more general than one comprising just dark matter (DM) 
as signaled by the presence of the time-dependent factor $C(\tau)$ in our Eq.~(\ref{eq:eom_v1}).
It can in fact describe a richer dynamic, with more degrees of freedom, to the extent that a quasi-static approximation is valid (see below as well as \cite{Sefusatti:2011cm} for an example of such a system).

For the purposes of this work, flat $\Lambda$CDM model is assumed as 
$\Omega_{\rm m}=0.27$, 
$\Omega_{\Lambda}=0.73$, $h=0.7$. For the primordial density power spectrum 
we use the BBKS \cite{Bardeen:1985tr} approximation for initial conditions.

\section{Equations of motion}
\label{sec:eom}
As we shall see, our results reduce, in the appropriate limit, to dark matter in a FLRW background \cite{Bernardeau:2001qr}. 
For such a dynamics the time-dependent quantity $C$ introduced below simplifies to $C(\tau)=1$,  we keep it to underscore the generality of our results. 
We describe our system as a fluid whose equations of motion (in the non-relativistic limit)  for the fluctuations of density contrast $\delta$ and peculiar velocity $v^i$ are:
\begin{align}
\label{eq:eom_v1}
&\frac{\p \delta_{\bf k}}{\p \tau} +C \theta_{\bf k}=-\int_{\bf{q}_{12}} \delta^{D}_{\bf{k}-\bf{q}_{12}} \alpha_{\bf{q}_1,\bf{q}_2} \theta_{\bf{q}_1}\delta_{\bf{q}_2} \\
&\frac{\p \theta_{\bf k}}{\p \tau} +\mathcal{H} \theta_{\bf k}+\frac{3}{2}\Omega_m\mathcal{H}^2 \delta_{\bf{k}} = -\int_{\bf{q}_{12}} \delta^{D}_{\bf{k}-\bf{q}_{12}}  
\beta_{\bf{q}_1,\bf{q}_2} \theta_{\bf{q}_1}\theta_{\bf{q}_2}\, , \nonumber
\end{align}
where we used the notation $ \delta^{D}_{\bf{k}-\bf{q}_{12}} =  \delta^{D}(\bf{k}-\bf{q}_{1}-\bf{q}_{2})$, as well as other
standard notation, such that $\theta=\p_i v^i$, the variable $\tau$ is conformal time, $\mathcal{H} = d \ln a/d \tau$, 
the kernels are $\alpha =1+({\bf q}_1\cdot {\bf q}_2)/{\bf q}_1^2$,
$\beta =({\bf q}_1+{\bf q}_2)^2 ({\bf q}_1\cdot{\bf q}_2) /2q_1^2 q_2^2$, 
and the Poisson equation reads  $\nabla^2 \Phi \propto \mathcal{H}^2\delta$, where $\Phi$ is the Newtonian potential. 
As mentioned, one of the systems whose dynamics is captured by Eq.~(\ref{eq:eom_v1}) is the clustering quintessence model in 
the vanishing sound speed ($c_s\rightarrow0$) limit \cite{Sefusatti:2011cm}.
In particular, the dictionary between the variables just above and the standard ones is:
\begin{align}
\label{dictionary}
\delta= \delta_m+\delta_Q\,\frac{\Omega_{Q}(\tau)}{\Omega_m(\tau)} \,; \quad 
C(\tau)=1+(1+w) \frac{\Omega_{Q}(\tau)}{\Omega_m(\tau)} ,
\end{align}
where $\delta_m$ and $\delta_Q$ are respectively the dark matter and quintessence density contrast, $w$ is the equation of state parameter,
assumed to be constant in time. The quantities $\Omega_m$ and $\Omega_Q$ are the density parameters; related conventions are discussed in appendix \ref{appA}. 
It is convenient at this stage to extract the linear time behaviour. To such end, one introduces the linear growth function $D$ via:
\bea
\delta^{(1)}_{\bf k}(\tau)\equiv D(\tau)\delta^{\rm in}_{\bf k}\; ,\; \quad  \theta^{(1)}_{\bf k}(\tau)\equiv -\mathcal{H}(\tau)\frac{f(\tau)}{C(\tau)} D(\tau) \delta^{\rm in}_{\bf k}\,,
\eea  
where $f^{+,-}\equiv d{\rm ln}\,D^{+,-}/d{\rm ln\,}a$ is the linear growth rate and its two modes correspond to the solutions of the second order equation for $D$. 
The quantity $a$ above is the scale factor and $\delta_{\bf k}^{\rm in}$ represents the initial value of the density constrast.
In what follows we will make extended use of the solutions for $D$ in various cosmologies, for explicit expressions we refer the reader to the classics \cite{Bernardeau:2001qr}.
To tackle the full non-linear case, we finally switch, for the velocity variable, to:
\begin{align}
\Theta_{\bf k}\equiv -\frac{C}{\mathcal{H}f_+}\theta\quad {\rm with}\quad \Theta^{(1)}_{\bf k}=D_+ \delta^{\rm in}_{\bf k}\, .
\end{align} 
We are then after the full non-linear solution to:
\bea
\frac{\p \delta_{\bf k}}{\p \eta} - \Theta_{\bf k}= \frac{\alpha(\bf{q}_1,\bf{q}_2)}{C} \Theta_{\bf{q}_1}\delta_{\bf{q}_2}\; , \qquad \qquad\qquad\quad\, \nonumber\\
\frac{\p \theta_{\bf k}}{\p \eta} -\Theta_{\bf k} -\frac{f_-}{f_+^2}(\Theta_{\bf k}-\delta_{\bf k}) =\frac{\beta(\bf{q}_1,\bf{q}_2)}{C} \Theta_{\bf{q}_1}\Theta_{\bf{q}_2}\; ,
\eea
where we introduced variable $\eta = \ln D_+$ and the integral over ${\bf q}_1,..,{\bf q}_n$ has been be omitted.
We solve the system perturbatively, employing the ansatz:
\begin{align}
\delta_{\bf k}(\eta)
&=\sum_{n=1}^{\infty} 
F^{s}_n({\bf q}_1..{\bf q}_n,\eta) D_+^{n}(\eta)\delta_{\bf q_1}^{\rm in}..\delta_{\bf q_n}^{\rm in},\;\; \nonumber\\
\Theta_{\bf k}(\eta)
&=\sum_{n=1}^{\infty} 
G^{s}_n({\bf q}_1..{\bf q}_n,\eta) D_+^{n}(\eta)\delta_{\bf q_1}^{\rm in}..\delta_{\bf q_n}^{\rm in}.
\end{align}

\newpage 

\section{Recursions \& kernel time dependence}
\label{sec:kernels}

\noi After symmetrization the $F$ and $G$ equations of motion are:
\begin{align}
&\dot{F}^{s}_{n}(\vec{q}_1,..,\vec{q}_n, \eta) + n\,{F}^{s}_{n} - {G}^{s}_{n} = \frac{1}{C(\eta)}  h^{(n)}_\alpha (\vec{q}_1,..,\vec{q}_n, \eta) \nonumber\\ 
&\dot{G}^{s}_{n}(\vec{q}_1,..,\vec{q}_n \eta) + (n-1)\,{G}^{s}_{n} \nonumber\\
&\hspace{1.4cm}-\frac{f_{-}}{f^2_{+}}\left( {G}^{s}_{n} - {F}^{s}_{n}  \right)= \frac{1}{C(\eta)} h^{(n)}_\beta (\vec{q}_1,..,\vec{q}_n, \eta) 
\label{eq:FsGs}
\end{align}
where we used the shorthand notation $\dot{ } = \frac{\partial}{\partial \eta} $, and introduced the source terms
\begin{align}
h^{(n)}_\alpha &(\vec{q}_1,..,\vec{q}_n, \eta) =  \sum_{\pi- {\rm all}} \sum^{n-1}_{m=1}\alpha({\vec{p}_m,\vec{p}_{n-m}}) \nonumber \\
&\hspace{1.6cm} \times G^S_m(\vec{q}_1,..,\vec{q}_m,\eta) F^S_{n-m}(\vec{q}_{m+1},..,\vec{q}_n,\eta) \nonumber \\
&= \sum^{n-1}_{m=1}\frac{m!(n-m)!}{n!} \sum_{\pi-{\rm cross}}\,\alpha({\vec{p}_{m_i}},{\vec{p}_{{m-n}_i}}) G^S_m F^S_{n-m} \nonumber \\
&= \left(\sigma(n)+\sum^{\left \lfloor (n-1)/2 \right \rfloor }_{m=1}\frac{m! (n-m)!}{n!}\right) \nonumber\\
&\hspace{0.1cm} \times \sum_{\pi-{\rm cross}}\,\Big[ \alpha({\vec{p}_{m_i}},{\vec{p}_{{m-n}_i}}) G^S_m F^S_{n-m} \nonumber\\
&\hspace{3.5cm} + \,\alpha({\vec{p}_{{m-n}_i}},{\vec{p}_{m_i}}) G^S_{n-m} F^S_m \Big]  \nonumber\\
h^{(n)}_\beta  &(\vec{q}_1,..,\vec{q}_n, \eta) = \sum_{\pi- {\rm all}} \sum^{n-1}_{m=1}\beta({\vec{p}_m},{\vec{p}_{m-n}}) \nonumber\\
&\hspace{1.7cm} \times G^S_m(\vec{q}_1,..,\vec{q}_m,\eta)  G^S_{n-m}(\vec{q}_{m+1},..,\vec{q}_n,\eta) \nonumber \\
&=2 \left(\sigma(n)+\sum^{\left \lfloor (n-1)/2 \right \rfloor }_{m=1}\frac{m! (n-m)!}{n!}\right) \nonumber\\
&\hspace{2.0cm} \times \sum_{\pi-{\rm cross}}\,\beta({\vec{p}_m,\vec{p}_{n-m}}) \, G^S_m G^S_{n-m}
\label{eq:hahb}
\end{align}
where $\sigma(n) = \left[1+(-1)^n\right] \frac{1}{4}\frac{(n/2!)^2}{ n!}$, and
where by \textit{cross} permutations it is meant those that exchange momenta in the $(1....m )$ set with those in the $(m+1...n)$ one. In the last line of Eq.~\eqref{eq:hahb} we have removed the double counting for $n$ even. The following quantities are also employed: $\vec{p}_m=\vec{q}_1+..+\vec{q}_m$; $\vec{p}_{n-m}=\vec{q}_{m+1}+..+\vec{q}_n$ and the index ``$i$" runs within cross permutations. Similarly, compact recursion relations in the EdS-like approximation \footnote{Strictly speaking in an EdS Universe the linear growth $D_+ = a$. The approximation commonly used in $\Lambda$CDM simply replaces $D_+$ where the scale factor would be in the exact EdS Universe. We shall mean precisely  this approximation when in the manuscript we refer to EdS-like approximation.} were recently derived in \cite{Bertolini:2015fya}. Henceforth we drop the symmetrization label, $s$, from $F$ and $G$  kernels.
Combining the two equations in Eq \eqref{eq:FsGs} one readily obtains for the first kernel:
\begin{align}
\ddot{F}_{n}+ &\dot{F}_{n} \left(2n-1 -\frac{f_{-}}{f_{+}^2}  \right)  +(n-1)F_{n}\left(n-\frac{f_{-}}{f_{+}^2}  \right)\nonumber \\ =& \frac{1}{C}\Bigg[h^{(n)}_{\beta}+
\left(n-1-\frac{\dot{C}}{C}  -\frac{f_{-}}{f_{+}^2}\right)h^{(n)}_{\alpha}  +\dot{h}^{(n)}_{\alpha}  \Bigg], 
\end{align}
whose solution reads:
\begin{align}
\label{Fnin}
&F_{n}(\eta)=\int_{-\infty}^{\eta} \frac{d\tilde{\eta}}{C(\tilde{\eta})}\left\{ e^{(n-1)(\tilde{\eta}-\eta)}\frac{\tilde{f}_{+}}{\tilde{f}_{+}-\tilde{f}_{-}} \times\right.  \\
&~~\left.\Bigg[\left(\tilde{h}^{(n)}_{\beta}-\frac{\tilde{f}_{-}}{\tilde{f}_{+}}\tilde{h}^{(n)}_{\alpha}  \right)+e^{\tilde{\eta}-\eta}\,\frac{D_{-}(\eta)}{\tilde{D}_{-}(\eta)} \left(\tilde{h}^{(n)}_{\alpha} -\tilde{h}^{(n)}_{\beta}\right)\Bigg] \right\}\, ,\nonumber
\end{align}
where in deriving the above we have used the equation of motion (e.o.m.) for the growing and decaying solutions for the linear growth factor $D$, $D_{+}(\eta), D_{-}(\eta)$. 
Using again the first equation in Eq.~\eqref{eq:FsGs} one immediately gets the solution for the $G$ kernels:
\begin{align}
\label{Gnin}
&G_{n}(\eta)=\int_{-\infty}^{\eta} \frac{d\tilde{\eta}}{C(\tilde{\eta})}\left\{ e^{(n-1)(\tilde{\eta}-\eta)}\frac{\tilde{f}_{+}}{\tilde{f}_{+}-\tilde{f}_{-}} \times\right. \\
&~\left.\Bigg[\left(\tilde{h}^{(n)}_{\beta}-\frac{\tilde{f}_{-}}{\tilde{f}_{+}}\tilde{h}^{(n)}_{\alpha}  \right)+e^{\tilde{\eta}-\eta}\frac{f_{-}}{f_{+}}\frac{D_{-}(\eta)}{\tilde{D}_{-}(\eta)} \left(\tilde{h}^{(n)}_{\alpha} -\tilde{h}^{(n)}_{\beta}\right)\Bigg] \right\}\, .\nonumber
\end{align}
Eqs.(\ref{Fnin}) and (\ref{Gnin}) are new integral solutions, to all orders, for the kernels $F$ and $G$ describing dark matter as well as more general setups. An integral solution limited to the quadratic case was found in \cite{Sefusatti:2011cm} (see also \cite{Anselmi:2011ef,D'Amico:2011pf} for related work), where also a differential ansatz for the quadratic solution was provided.
Our results for $\delta^{(2)}, \Theta^{(2)}$ agree with \cite{Sefusatti:2011cm} for an Einstein-de Sitter background.

It is very useful, especially for computational purposes, to also provide a general differential ansatz for $\delta^{(n)}, \Theta^{(n)}$. We briefly review the results for the second order fields (due to the work in \cite{Sefusatti:2011cm}) and then present for the 
first time the results for third order fields $\delta^{(3)}, \Theta^{(3)}$.
Following the notation in \cite{Sefusatti:2011cm} second order kernels are given by
\begin{align}
F_{2}(\eta,{\bf q}_1,{\bf q}_2)&= -\frac{1}{2} \left [ 1 - \epsilon- \frac{3}{2}\nu_2 \right ]\alpha_s + \frac{3}{2} \left [ 1 - \epsilon - \frac{1}{2}\nu_2 \right] \beta \nonumber \\
G_{2}(\eta,{\bf q}_1,{\bf q}_2)&= -\frac{1}{2} \left [ 1 - \epsilon - \frac{3}{2}\mu_2 \right ]\alpha_s + \frac{3}{2} \left [ 1 - \epsilon - \frac{1}{2}\mu_2 \right] \beta\, ,
\label{eq:diff_an_2}
\end{align}
where for simplicity we have suppressed the time dependence in $\epsilon, \mu_2$ and $\nu_2$ as well as the momenta dependence in $\alpha$ and $\beta$. The definition of the function $\epsilon$ is chosen according to $\epsilon(\eta)=1-e^{-\eta} \int^{\eta}_{-\infty}d\tilde{\eta}\[e^{\tilde{\eta}}/C(\tilde{\eta})\]$, so that it vanishes in the simplifying case where $C=1$. We shall see the e.o.m.s satisfied by $\mu_n, \nu_n$ in what follows.
We provide here the formal ansatz for the third order kernels and then proceed to write more explicitly their respective building blocks:
\begin{align}
F_{3}(\eta,{\bf q}_1,{\bf q}_2,{\bf q}_3)&= (1-\epsilon^{(2)}) \mathcal{F}_3^{\epsilon}+\nu_3 \mathcal{F}_3^{\nu_3}+  (1-\epsilon^{(1)}) \nu_2 \mathcal{F}_3^{\nu_2} \nonumber\\
&\hspace{2.cm} + \lambda_1 \mathcal{F}_3^{\lambda_1} +\lambda_2 \mathcal{F}_3^{\lambda_2}\,\,\, \nonumber \\
G_{3}(\eta,{\bf q}_1,{\bf q}_2,{\bf q}_3)&= (1-\epsilon^{(2)}) \mathcal{G}_3^{\epsilon}+\mu_3 \mathcal{G}_3^{\mu_3}+  (1-\epsilon^{(1)}) \mu_2 \mathcal{G}_3^{\mu_2} \nonumber\\
&\hspace{2.cm} + \kappa_1 \mathcal{G}_3^{\kappa_1} +\kappa_2 \mathcal{G}_3^{\kappa_2}  \, .
\label{eq:diff_an_3}
\end{align}
The functions above of the type $\mathcal{F}^{}$ and $\mathcal{G}^{}$ are the third order counterpart of the $\alpha$ and $\beta$ expressions above and, as we shall see in detail, depend only on momenta, e.g. $\mathcal{F}^{}=\mathcal{F}^{}({\bf q}_1,{\bf q}_2,{\bf q}_3)$. The quantities  $\nu_n, \mu_n$  are time-only dependent variables defined (see \cite{Bernardeau:2001qr} and also \cite{Bernardeau:1993qu}) as the angle average of $F_{n}, G_{n}$ weighted by $n!$ and we have introduced above the function $\epsilon^{(2)}$:
\begin{align}
\epsilon^{(2)}= 2\int_{-\infty}^{\eta}d\tilde{\eta} ~e^{2(\tilde{\eta}-\eta)} \left( 1-\frac{1-\epsilon}{C(\tilde{\eta})} \right),
\label{eq:eps_2}
\end{align}
which is a generalisation of the $\epsilon$ function defined above and, similarly to $\epsilon$, vanishes in the $C=1$ limit.
The angle-averaged kernel dynamics is governed by the following equations \cite{Bernardeau:2001qr}:
\begin{align}
\label{eq:average}
&\dot{\nu}_n + n\,\nu_n-\mu_n = \frac{1}{C} \sum_{m=1}^{n-1} {n \choose m} \mu_m \, \nu_{n-m}\, , \\
&\dot{\mu}_n + (n-1)\mu_n -\tfrac{f_-}{f_+^2}(\mu_n-\nu_n) = \frac{1}{3C} \sum_{m=1}^{n-1} {n \choose m} \mu_m \, \mu_{n-m}\,,\nonumber
\end{align}
with ``initial conditions" $\nu_1=\mu_1=1$. Upon implementing the ansatz from Eq.~\eqref{eq:diff_an_3} in Eq.~(\ref{eq:FsGs}) 
and using Eq.~(\ref{eq:diff_an_2}) and Eq.~(\ref{eq:eps_2}) one derives the relations 
$\mathcal{F}_3^{\epsilon}=\mathcal{G}_3^{\epsilon},\, \mathcal{F}_3^{\nu_3} =\mathcal{G}_3^{\mu_3},\,  \mathcal{F}_3^{\nu_2}=\mathcal{G}_3^{\mu_2}\,,  
\mathcal{F}_3^{\lambda_1}=\mathcal{G}_3^{\kappa_1}\,, \mathcal{F}_3^{\lambda_2}=\mathcal{G}_3^{\kappa_2}  \,$.
The repeated use of Eq.~(\ref{eq:average}) leads to the momentum dependence of the functions 
$ \mathcal{F}^{\epsilon}_3, ~\mathcal{F}^{\nu_2}_3,~\mathcal{F}^{\nu_3}_3,~\mathcal{F}_3^{\lambda_1}$ and $\mathcal{F}_3^{\lambda_2}$,
which can then be read off from that of known function in Eq.~(\ref{eq:FsGs}).
Following the same procedure, we obtain four new momentum-independent differential 
equations for $\lambda_i$ and $\kappa_i$
\begin{align}
&\dot{\lambda_i}+3 \lambda_i-\kappa_i=\frac{1}{C}\left(\nu_2\, c_{\lambda_i}^{\nu2}+\mu_2\, c_{\lambda_i}^{\mu2}\right)\; , \nonumber\\
&\dot{\kappa_i}+2 \kappa_i-\frac{f_-}{f^2_+}(\kappa_i-\lambda_i)= \frac{1}{C}\,\mu_2\, c_{\kappa_i}^{\mu2}\; ,
\label{eq:lambda_kappa}
\end{align}
where index $i$ can take values $\left\{1,2 \right\}$.
Note that these equations are of the same form as equations for $\nu_3$ and $\mu_3$ in Eq.~\eqref{eq:average}
and can be very efficiently integrated numerically. 
In order to match the results to the initial conditions of the EdS type  we choose for the RHS parameters
in Eq.~\eqref{eq:lambda_kappa} above:
\be
c_{\lambda_1}^{\nu2}=c_{\lambda_1}^{\mu2}=c_{\lambda_2}^{\nu2}=2 c_{\lambda_2}^{\mu2}=1\; , \qquad \; 
c_{\kappa_1}^{\mu2}=c_{\kappa_2}^{\mu2} = 0 \; .
\label{eq:c_choice}
\ee
With this choice we derive the momentum dependence of the third order kernels:
\begin{align}
\mathcal{F}_3^{\epsilon}&=-\frac{1}{12}\Bigg[(\alpha^{s}_{12,3}-3\beta_{12,3})(3\beta_{12}-\alpha^{s}_{12}) +  {\rm 2\, perm.}_{\rm cross}\Bigg]\; , \quad\;\;\nonumber\\
\mathcal{F}_3^{\nu_3}&=\frac{1}{8}\Bigg[\left(\, \alpha^s_{1,23}(\alpha^s_{23}-3\beta_{23})\right.  \nonumber \\ 
                               &\hspace{1.2cm}\left.+\beta_{1,23}(\alpha^s_{23}+\beta_{23})\,\right)+   {\rm 2\, perm.}_{\rm cross}\Bigg]\; , \nonumber
\end{align}
\begin{align} 
\mathcal{F}_3^{\nu_2}&=\frac{1}{4}\Bigg[(\alpha^{s}_{12,3}-\beta_{12,3})(3\beta_{12}-\alpha^{s}_{12}) +  {\rm 2\, perm.}_{\rm cross}\Bigg]\; , \nonumber\\
\mathcal{F}_3^{\lambda_1}&=\frac{1}{16}\Bigg[\left( \,\alpha_{12,3}(3\alpha^s_{12}+7\beta_{12}) + \alpha_{1,23}(-9\alpha^s_{23}+19\beta_{23})\right. \nonumber \\   
                                        &\hspace{0.9cm}-2 \left. \beta_{1,23}(\alpha^s_{23}+9\beta_{23})\, \right) +   {\rm 2\, perm.}_{\rm cross}\Bigg]\; , \nonumber\\
\mathcal{F}_3^{\lambda_2}&=\frac{1}{4}\Bigg[ \left(\, \alpha_{1,23}(3\alpha^s_{23}-5\beta_{23}) - \alpha_{12,3}(\alpha^s_{12}+\beta_{12})\right. \nonumber \\
                                        &\hspace{0.9cm}-2 \left. \beta_{1,23}(\alpha^s_{23}-3\beta_{23}) \,\right) +   {\rm 2\, perm.}_{\rm cross}\Bigg]\; ,
\end{align}
\noi where again by \textit{cross} it is meant the permutations that exchange momenta in the $(1....m )$ set with those in the $(m+1...3)$ one.
\section{Results for one-loop power spectrum}
\label{sec:one_loop}
\begin{figure*}[t!]
\includegraphics[scale=0.66]{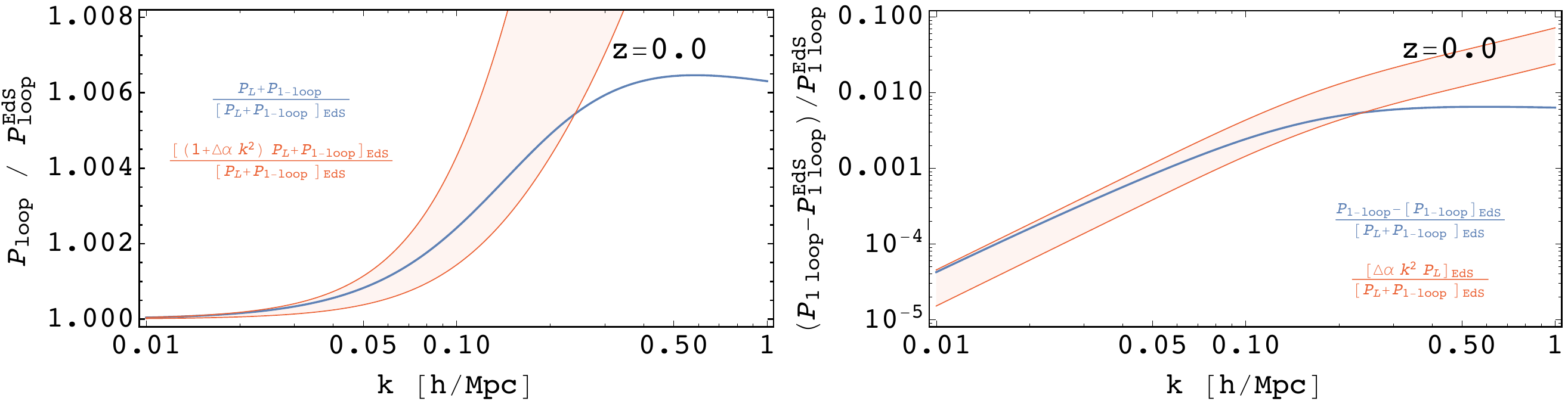}
\caption{\textit{Left}: the ratio between our density power spectrum result and the EdS-extended (see Ref.[41])  power spectrum. \textit{Right:} the difference between the 1-loop terms from the same quantities normalized by the full EdS-extended power spectrum. The red band is obtained by means of the usual $ \Delta \alpha\, k^2P_{\rm lin}$ approximation.} 
\label{fig:p00}
\end{figure*}

Using the kernels derived in section \ref{sec:kernels} we can 
proceed to an illustrative example and compute the power spectrum at  one loop:
\be
P_{\rm 1-loop}(k,a) = P_{\rm L}(k,a) + P_{22}(k,a)+2P_{13}(k,a) + P_{\rm c.t.}(k,a),
\ee
where individual contributions are given as
\begin{align}
P_{{\rm L},k}(a) &= D^2_+(a) P^{\rm in}_{\bf k},\; \nonumber\\
P_{22,k}(a) &= 2D^4_+(a) \int_{\bf q} \big[ F_{2}({\bf k} -{\bf q}, {\bf q},a) \big]^2 P^{\rm in}_{|{\bf k} - {\bf q}|} P^{\rm in}_{\bf q} \, , \nonumber\\
P_{13,k}(a) &= 3D^4_+(a) P^{\rm in}_{\bf k} \int_{\bf q} F_{3}({\bf k}, -{\bf q}, {\bf q},a)  P^{\rm in}_{\bf q}\, ,
\end{align}
 and $P^{\rm in}_{k}$ stands for the initial time-independent power spectrum. It is understood that the short-scale dynamics will be encoded in the appropriate counter-term part ($P_{\rm c.t.}$), whose numerical value is to be determined via e.g. N-body simulations. Due to rotational invariance, this is expected to be of the type $\propto k^2/k^2_{\rm NL}\,P_{\rm L}$ \cite{Carrasco:2012cv}.

Let us proceed with the detailed calculation of $P_{22}$ and $P_{13}$ term. The goal is to separate the time dependence and momentum dependence since this form enables 
practical evaluation of the contributing terms. Using Eq.~\eqref{eq:diff_an_2} for $F_2$
kernels one obtains for the $P_{22}$:
\begin{align}
\frac{P_{22,k}}{D_+^4} &= \left [ 1 - \epsilon- \tfrac{3}{2}\nu_2 \right ]^2\mathcal{I}^{\alpha}_{22,k} + \left [ 1 - \epsilon - \tfrac{1}{2}\nu_2 \right] ^2 \mathcal{I}^{\beta}_{22,k}  \nonumber \\
&\hspace{0.5cm} - \left [ 1 - \epsilon- \tfrac{3}{2}\nu_2 \right ] \left [ 1 - \epsilon - \tfrac{1}{2}\nu_2 \right]  \mathcal{I}^{\alpha\beta}_{22,k}  \, , \nonumber\\
&= (1 - \epsilon)^2 ~ \Big( \mathcal{I}^{\alpha}_{22,k} - \mathcal{I}^{\alpha\beta}_{22,k} + \mathcal{I}^{\beta}_{22,k} \Big) \nonumber \\
&\hspace{0.5cm} + (1 - \epsilon) \nu_2 ~ \Big( 3 \mathcal{I}^{\alpha}_{22,k} - 2 \mathcal{I}^{\alpha\beta}_{22,k} + \mathcal{I}^{\beta}_{22,k} \Big) \nonumber\\
&\hspace{0.5cm} + \nu_2^2 ~ \frac{1}{4} \Big( 9 \mathcal{I}^{\alpha}_{22,k} - 3 \mathcal{I}^{\alpha\beta}_{22,k} + \mathcal{I}^{\beta}_{22,k} \Big)
\end{align}
where the time-independent contributions used above are:
\begin{align}
\mathcal{I}^{\alpha}_{22,k}\equiv&~ \frac{1}{2}\int_{\bf q} \, \big[ \alpha_s({\bf k}-{\bf q},{\bf q}) \big]^2 P^{\rm in}_{|{\bf k} - {\bf q}|} P^{\rm in}_{\bf q}\, , \nonumber \\
\mathcal{I}^{\beta}_{22,k}\equiv& ~ \frac{9}{2}\int_{\bf q}\, \big[ \beta({\bf k}-{\bf q},{\bf q}) \big]^2 P^{\rm in}_{|{\bf k} - {\bf q}|} P^{\rm in}_{\bf q}\, ,\nonumber  \\
\mathcal{I}^{\alpha\beta}_{22,k} \equiv& ~ 3 \int_{\bf q}\, \alpha_s({\bf k}-{\bf q},{\bf q})\beta({\bf k}-{\bf q},{\bf q}) P^{\rm in}_{|{\bf k} - {\bf q}|} P^{\rm in}_{\bf q} \, .
\end{align}
Similarly, using Eq.~\eqref{eq:diff_an_3} for $F_3$ one readily obtains
expression for $P_{13}$, which can be organised as follows:
\begin{align}
\frac{P_{13,k}}{D_+^4} &= (1-\epsilon^{(2)})~\mathcal{I}^{\epsilon}_{13,k} + \nu_3 ~ \mathcal{I}^{\nu_3}_{13,k} +(1-\epsilon) \nu_2~ \mathcal{I}^{\nu_2}_{13,k}  \nonumber \\
& \hspace{0.5cm} + \lambda_1\mathcal{I}^{\lambda_1}_{13,k} + \lambda_2 \mathcal{I}^{\lambda_2}_{13,k} \, , 
\end{align}
where we have again isolated the time independent contributions:
\begin{align}
\mathcal{I}^{\epsilon}_{13,k} \equiv & ~ 3 P^{\rm in}_{\bf k} \int_{\bf q} \, \mathcal{F}_3^{\epsilon}({\bf k},-{\bf q},{\bf q}) P^{\rm in}_{\bf q}\, , \nonumber\\
\mathcal{I}^{\nu_3}_{13,k} \equiv & ~ 3 P^{\rm in}_{\bf k} \int_{\bf q} \, \mathcal{F}_3^{\nu_3}({\bf k},-{\bf q},{\bf q}) P^{\rm in}_{\bf q}\, , \nonumber\\
\mathcal{I}^{\nu_2}_{13,k}  \equiv & ~ 3 P^{\rm in}_{\bf k} \int_{\bf q} \, \mathcal{F}_3^{\nu_2}({\bf k},-{\bf q},{\bf q}) P^{\rm in}_{\bf q}\, , \nonumber\\
\mathcal{I}^{\lambda_1}_{13,k} \equiv & ~ 3 P^{\rm in}_{\bf k} \int_{\bf q} \, \mathcal{F}_3^{\lambda_1}({\bf k},-{\bf q},{\bf q}) P^{\rm in}_{\bf q}\, ,\nonumber\\
\mathcal{I}^{\lambda_2}_{13,k} \equiv & ~ 3 P^{\rm in}_{\bf k} \int_{\bf q} \, \mathcal{F}_3^{\lambda_2}({\bf k},-{\bf q},{\bf q}) P^{\rm in}_{\bf q}\, .
\end{align}

\begin{figure*}[t!]
\includegraphics[scale=0.66]{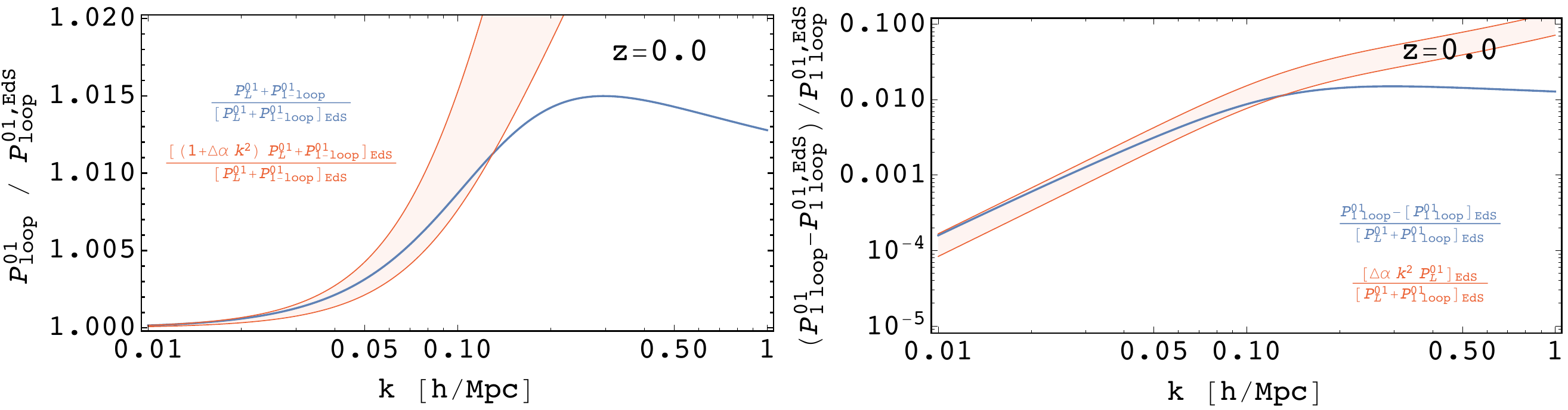}
\caption{\textit{Left}: the ratio between our density-momentum power spectrum result, $P_{01}$, and the EdS-extended (see Ref.[41])  power spectrum. \textit{Right:} the 
difference between the 1-loop terms from the same quantities normalized by the full EdS-extended power spectrum. 
The red band is obtained by means of the usual $ \Delta \alpha \, k^2P_{\rm lin}$ approximation. For the sake of comparison, we note that the $y$-axis in the left panel here is different with respect to the one in Fig.~\ref{fig:p00}. } 
\label{fig:p01}
\end{figure*}

We note in passing that in our case for $C\not=1$ one does not necessarily expect the cancellation \cite{Jain:1995kx,Peloso:2013zw,Kehagias:2013yd} between $P_{22}$ and  $P_{13}$ 
in the IR as the initial conditions do not \cite{Bernardeau:2011vy} always conform to the usual expressions. 
This may pave the way to interesting observational consequences. In Fig. \ref{fig:p00}, both the ratio between our density power spectrum result and the EdS-extended (see Ref.[41]) one (\textit{left}) and the difference between the 1-loop terms of the same quantities normalized by the full power spectrum (\textit{right}) are presented as a function of $k$. For $k$'s into the quasi-linear regime we find a difference close to 1\%. Note that the red band in both Fig.~\ref{fig:p00} and Fig.~\ref{fig:p01} is obtained by means of the usual $\propto k^2 P_{\rm lin}$ approximation and serves as a measure of the uncertainty at the one-loop perturbative order. Incidentally, it can be also thought of as a proxy for the 1-loop counterterm contribution.

It is also instructive to look at the cross power spectrum of the density and momentum fields. 
The continuity equation relates the scalar component of the momentum field and the time derivative of the density field $\tfrac{d}{d\tau} \delta - i k p_s = 0$. This gives us the simple 
relation between the density-momentum power spectrum and the time derivative of the 
density power spectrum $P_{01} = i/k P_{\delta\delta'} = i/2k \tfrac{d}{d\tau} P_{\delta\delta}$ 
(see e.g. \cite{Seljak:2011tx,Vlah:2012ni}). 
In Fig. \ref{fig:p01} we show the ratio between our density-momentum power spectrum $P_{01}$ and 
the EdS-extended one. We note that the effects due to the exact time evolution are more noticeable here, reaching 1.5\% in the mildly nonlinear regime. Therefore, these effects seem to be of importance if percent precision is to be reached, especially for observables in redshift space (see \cite{Seljak:2011tx, Okumura:2011pb,Vlah:2012ni, Vlah:2013lia, Okumura:2013zva} for the relation between velocity momentum statistics and redshift space observables).
We note that these effects might be larger for higher order velocity momentum statistics.

\section{Conclusions}
\label{sec:conclusions}

We have derived the all-order perturbative solution to a LSS dynamical system that goes beyond $\Lambda$CDM. 
As an application of our results, we provided the 1-loop calculation for the density and density-momentum power spectrum. The difference with respect to the standard approximation (see Ref[41]) is, in the quasi-linear regime, close to 1\% for the density power spectrum and over 1\% for the density momentum power spectrum.

\noindent \acknowledgements We are delighted to thank Massimo Pietroni for illuminating discussions. We are also indebted to Emiliano Sefusatti and Uros Seljak for insightful comments.
M.F. is supported in part by NSF grant PHY-1068380 and Z.V. is supported in part by the U.S. Department of Energy contract to SLAC no. DE-AC02-76SF00515.

\appendix
\section{Review of linear growth results and conventions}
\label{appA}

Combining the continuity and Euler equation \eqref{eq:eom_v1}, we have a second order differential 
equation for the density
\be
\frac{d^2 \delta^{(1)}_{\bf{k}}(\tau)}{d \tau^2} + \left(\mathcal{H} - \frac{d \ln C}{d \tau} \right) \frac{d \delta^{(1)}_{\bf{k}}(\tau)}{d \tau} 
- \frac{3}{2}\Omega_MC\mathcal{H}^2 \delta^{(1)}_{\bf{k}} (\tau) = 0  \, .
\ee
Combining this equation with the Friedman equations:
\begin{align}
3 \mathcal{H}^2 &= 8 \pi G a^2 (\bar \rho_M + \bar \rho_Q) , \\
\frac{d \mathcal{H}}{d \tau} &= - \frac{4\pi G}{3} a^2 (\bar \rho + 3 \bar p ) = - \frac{4\pi G}{3} a^2 (\bar \rho_M + (1+3w) \bar \rho_Q ) \, , \nonumber
\end{align}
using the definitions $\Omega_\alpha = \bar \rho_\alpha / (\bar \rho_M + \bar \rho_Q) = 8 \pi G H_0^2 a^2 \bar \rho_\alpha / (3 \mathcal{H}^2)$ we have
\begin{align}
\frac{d \mathcal{H}}{d \tau} &= -  \frac{1}{2} \mathcal{H}^2 \big( \Omega_M + (1+3w) \Omega_Q \big) = -  \frac{1}{2} \mathcal{H}^2 \big( 1 + 3w \Omega_Q \big) \, , \nonumber\\
\Omega_M &+ \Omega_Q = 1 \, , 
\end{align}
and from the first Friedman equation we get explicitly the evolution of the
Hubble parameter in $w$CDM ($w$ is constant) universe,
\be
H(a) = H_0 \sqrt{\Omega_{M,0} a^{-3} + \Omega_{Q,0} a^{-3(1+w)}}  \, .
\ee
Combining all of the above and changing the variables, we can rewrite the linear density equation:
\begin{align}
\frac{d^2 \delta^{(1)}_{\bf{k}}(a)}{d \ln a^2} + &\left( \frac{1}{2} (1-3w\Omega_Q)  \right. \\
&\hspace{0.9cm} \left. - \frac{d \ln C}{d \ln a} \right) \frac{d \delta^{(1)}_{\bf{k}}(a)}{d \ln a} 
- \frac{3}{2}\Omega_M \delta^{(1)}_{\bf{k}} (a) = 0 \, . \nonumber
\end{align}
The solutions to this equation can formally be written as 
\begin{align}
 \delta^{(1)}_{\bf k }(a) = D^+(a)C_+({\bf k})  + D^-(a)C_-({\bf k }) \, ,
\end{align}\\
where we have introduced the linear growth factor $D^{+/-}$ describing the growing and decaying modes. 
Explicit expressions can be found for many cosmologies (see e.g. \cite{Bernardeau:2001qr}).
For numerical evaluation it is convenient to rewrite the Growth equations in the form
\be
a^2 \frac{d^2 D(a)}{da^2} + a F(a)\frac{d D(a)}{da} - \frac{3}{2} \Omega_M C D(a) = 0  \, ,
\ee
where one introduces a function $F(a) \equiv 3/2 (1- w \Omega_Q) - d\ln C/d\ln a$.
There are two conventions in the literature for normalizing a growing mode. One normalization convention stems from requiring that a growing mode is equal to the scale factor in the matter dominated epoch: 
$D+(a)_{\rm EdS} = a$.
Here, EdS stands for the `Einstein de-Sitter' Universe which is a flat, matter dominated universe. 
Explicitly, the solutions for a $w$CDM Universe are
\begin{align}
D^+ (a) &= \frac{5}{2} H_0^2 \Omega_{M,0} H(a) \int_0^a \frac{C(\tilde a) d \tilde a }{[\tilde a H(\tilde a)]^3}  \, , \nonumber \\
D^{-}(a) &= H(a)  \, .
\end{align}

\vfill
\bibliographystyle{arxiv_physrev}
\bibliography{ms}

\end{document}